\documentclass[12pt,a4paper,english,superscriptaddress,aps,nofootinbib]{revtex4}
\usepackage[utf8]{inputenc}
\usepackage[T1]{fontenc}
\usepackage{amsmath,amssymb,graphicx}
\makeatletter
\usepackage{babel}
\usepackage[active]{srcltx}
\usepackage{graphicx,color}
\usepackage{changebar}
\usepackage{hyperref}
\usepackage[T1]{fontenc}
\usepackage{esint}
\usepackage{multirow}
\usepackage{dsfont}
\usepackage{ae}
\usepackage{amsmath}
\usepackage{braket}
\usepackage{mathtools}
\usepackage{slashed}
\usepackage{empheq}
\usepackage{multirow}
\usepackage{graphicx}
\usepackage{amsfonts}
\usepackage{amsmath}
\newcommand{\be}{\begin{equation}}
	\newcommand{\ee}{\end{equation}}
\newcommand{\bea}{\begin{eqnarray}}
	\newcommand{\eea}{\end{eqnarray}}

\begin{document}
	\title{Properties of black hole vortex in Einstein's gravity}
	
\author{F. C. E. Lima}
\email{cleiton.estevao@fisica.ufc.br}
\affiliation{Universidade Federal do Cear\'{a}, Departamento do F\'{i}sica, Fortaleza, CE, 60455-760, Brazil.}

\author{A. R. P. Moreira}
\affiliation{Universidade Federal do Cear\'{a}, Departamento do F\'{i}sica, Fortaleza, CE, 60455-760, Brazil.}
 
\author{C. A. S. Almeida}
\email{carlos@fisica.ufc.br}
\affiliation{Universidade Federal do Cear\'{a}, Departamento do F\'{i}sica, Fortaleza, CE, 60455-760, Brazil.}

\begin{abstract}\vspace{0.4cm}
\noindent \textbf{Abstract:} We investigate the influence of the matter field and the gauge field on the metric functions of the AdS$_3$ spacetime of the Maxwell-Higgs model. By considering a matter field with a solitonic profile with the ability to adjust the field variable from kink to compact-like configurations, the appearance of black hole solutions is noticed for an event horizon at $r_{+} \approx 1.5$. An interesting result is displayed when analyzing the influence of matter field compactification on the metric functions. As we obtain compact-like field configurations the metric functions tend to a ``linearized behavior''. However, the compactification of the field does not change the structure of the horizon of the magnetic black hole vortex. With the ADM formalism, the mass of the black hole vortex is calculated, and its numerical results are presented. By analyzing the so-called ADM mass, it is observed that the mass of the black hole vortex increases as the cosmological constant becomes more negative, and this coincides with the vortex core becoming smaller. Nonetheless, this mass tends to decrease as the solitonic profile of the matter field becomes more compacted. Then, the black hole temperature study is performed using the tunneling formalism. In this case, it is perceived that the cosmological constant, and the $\alpha$-parameter, will influence the Bekenstein-Hawking temperature. In other words, the temperature of the structure increases as these parameters increase.

\end{abstract}
\maketitle

\thispagestyle{empty}

\section{Introduction}

Since the work of Abrikosov \cite{Abrikosov}, in condensed matter physics, and Nilsen-Olesen \cite{Nielsen} in field theory, on the existence of purely magnetic vortices, non-perturbative solutions have been a theoretical question of lasting interest \cite{KLee,Casana,LA,LPA}. These vortex structures have been intensively studied in flat \cite{LPA,Adam} and curved spacetime \cite{Kim,LA1}. In large part, this is due to its applications. For example, these structures can be related to anyons physics and the fractional Hall effect \cite{Ezawa}, as well as the physics of axions and the fractional angular momentum \cite{Nogueira}.

Nielsen and Olesen \cite{Nielsen} in their seminal work, investigated the vortices of an Abelian Maxwell-Higgs model. For this, they realized that the gauge field should be coupled minimally to a charged scalar field. Based on Ref. \cite{Nielsen}, several works with the proposal to investigate these structures \cite{Lima,LPA1,Han,Huang}, and their properties \cite{Ramadhan,DBazeia,DBazeia1}, have emerged in literature.

In general, the topological field solutions that describe the vortices are described by static solutions of nonlinear fields, which allow the spontaneous symmetry breaking mechanism \cite{Manton}. In this sense, such structures have been intensively studied, and their consequences are applied in several areas of interest, mainly in cosmology, where it is known that they are formed during the phase transitions \cite{Vilenkin}.

Generically, the vortices are structures that carry a quantized magnetic flux but no electrical charge. The vortices emerge for the first time in the condensed matter physics context to explain superconductivity \cite{Abrikosov,Albert}. However, soon after, these structures also were considered in the description of vortex lines for cosmic string, see Ref. \cite{Nielsen,Albert}. Recently, vortices have gained a reinterpretation in the cosmological scenario, or precisely, in the physical description of black holes \cite{GDvali}. In this scenario, vortices are structures that can be emitted from black holes and trap a quantized magnetic flux. Thereby, we call these configurations of black hole vortices. In this way, we base the interpretation of the black hole vortex on the description of the graviton condensate of a black hole and the correspondence between black holes and generic objects with maximum entropy compatible with unitarity, the so-called saturons \cite{GDvali}. Thus, this theoretical approach (or interpretation) can provide a topological explanation for the stability of extreme black holes and their repercussions on Hawking evaporation \cite{GDvali}.

An astrophysical object of great interest in contemporary physics is the black hole \cite{Frolov}. With the experimental evidence, researchers have become adept at studying these objects \cite{Akiyama}. These cosmological systems emerge as a solution to the equation of the gravitational field of Einstein \cite{Rindler}. The first known black hole solution was demonstrated by K. Schwarzschild, and describes the geometry of a spherical, static black hole. Then, Oppenheimer and Snyder showed that the Schwarzschild solution describes the final state of collapse of a massive, spherical star \cite{opp}.

One of the features of black hole solutions in any dimension of spacetime is their uniqueness. In general, these solutions are characterized by several parameters from asymptotic charges associated with global symmetries of the solution \cite{Cardoni}. More recently, an issue that has been highlighted is related to the singularity of black hole solutions in the presence of scalar fields \cite{Cardoni}. This question is related to the precise formulation of the hairless conjecture, i. e., the absence of non-trivial scalar field configurations in black hole background \cite{Bekenstein1,Bekenstein2,Teitelboim}. In this context, the relevant question concerns not only the presence of scalar hairs but also the possibility of having scalar charges independent of the black hole mass \cite{Cardoni}.

Although the non-perturbative topological vortex solutions have already been intensively studied, the study of the interaction effects of these structures with gravity is still poorly understood. This is because, in flat spacetime, Einstein's gravity is trivial in the sense that outside of localized sources, spacetime in a vacuum is locally flat \cite{Edery,Deser}. However, some interesting papers emerged from time to time. We can remark the BTZ black hole solution on an AdS$_3$ background, which is an interesting work in the (2+1)-dimensional Einstein gravity \cite{Banados,Banados1}. On the other hand, Cardoni \textit{et. al. } \cite{Cardoni} studied the analytical solutions of spherical scalar haired black holes in Einstein gravity on an AdS$_3$ background. In addition to a real scalar field, they also considered a complex scalar field with a potential, which allowed them to build vortex solutions in black hole.

Nowadays, the idea of a black hole is commonly linked to the concept of Hawking radiation, which makes the thermodynamic description of black holes a more accessible concept \cite{Gomes2018oyd}. As a matter of fact, the thermodynamics properties of black holes were intensively discussed by Bardeen \textit{et. al.} \cite{Bardeen}. In that work, they formulated the four laws of black hole thermodynamics, based on the ideas of Bekenstein \cite{Bekenstein3}. These laws were proposed due to their similarity to some mechanical properties of black hole, i. e., its mass, horizon area, and surface gravity. The fact is that today black holes can be considered thermal systems, and the laws that govern these objects are known as the thermodynamic laws of black holes.

Our work is motivated to understand how the topological vortex structures of Maxwell-Higgs Abelian model, can change the geometry of spacetime in (2+1)-dimensional Einstein gravity. We seek to answer whether it is possible for black holes to appear in a curved 3D spacetime in Einstein's gravity. Furthermore, we study what influence the matter field can have on the black hole if it becomes more compacted. For this, we start from a general metric, and consider a matter field with an axially symmetric solitonic profile.

This work is organized as follows. In Sec. II, we build our model and discuss the gauge field and the metric functions generated by the solitonic matter field.  Also is shown that the metric functions obtained describe a black hole type structure.  Finally, the behavior of the curvature scalar, the Kretschmann scalar, and the Ricci quadratic invariant are displayed. In Sec. II, the ADM mass of the black hole vortex is discussed, and its numerical results are presented. In Sec. IV,  the thermodynamics of the black hole is studied. Finally, in Sec. V, we discussed our findings.   

\section{The black hole vortex}
\label{sec2}
Since the seminal paper by Deser et al. \cite{Deser} and Witten \cite{Witten,Witten1}, relativistic problems in (2+1)-dimensional have become increasingly popular in the study of classical and quantum gravity theories \cite{Carlip}. To a certain degree, this is because, in $(2+1)$-dimensional models, the Einstein theory has the classical limit, i. e., the Newtonian limit and degrees of freedom of propagation \cite{Carlip}. Although, the four-dimensional gravitational models are more realistic. On the other hand, three-dimensional models are a laboratory that helps us to interpret higher-dimensional theories \cite{Carlip,Carlip1}. In 1992, Banados et al. affirmed that gravity in (2+1)-dimensional can have a black hole solution. In fact, these structures proposed by Banados are asymptotically anti-de Sitter without curvature singularity at the origin. However, the (2+1)-dimensional black hole structure has an event horizon that appears as the final state of collapsing matter \cite{Banados}. This event horizon resembles the event horizon of three-dimensional black holes.

Thinking about this, let us start our discussion by considering the Lagrangian density for the Maxwell-Higgs vortex coupled with Einstein gravity being
\begin{align}\label{eq1}
    \mathcal{L}=\sqrt{-g}\bigg[\frac{1}{16\pi G}(R-2\Lambda)+\frac{1}{2}(D_{\mu}\phi)^{\dagger}D^{\mu}\phi-\frac{1}{4}F_{\mu\nu}F^{\mu\nu}-\frac{\lambda}{4}(\vert\phi\vert^2-\nu^2)^2\bigg],
\end{align}
where $G$ is the Newton constant, $R$ is the Ricci scalar, $\Lambda$ is the cosmological constant, $\nu$ is the VEV of the theory, $F_{\mu\nu}$ is the electromagnetic tensor, $A_\mu$ is the gauge field, and the covariant derivative is defined as
\begin{align}\label{eq2}
    D_{\mu}\phi=\partial_\mu\phi+ieA_{\mu}\phi.
\end{align}

    Here we highlight that it is difficult to describe how to go from kinks to compact configurations in different scenarios. Discussions of compact solutions in models with standard dynamics were found in flat spacetime \cite{Bazeia}, which encouraged us to search for the influence of these structures in curved spacetime. To turn a kink configuration into compact-like configurations\footnote{The term compact-like refers to solitonic field configurations with a finite wavelength \cite{LGA}. Generally, one uses compact-like structures in describing topological structures associated with particles and cosmological objects, see Refs. \cite{Rosenau,LGA}.} it is necessary to write a potential that controls the fluctuations of the solutions. In a 2D model, the potential that controls the fluctuations of the kink mass increases to increasingly larger values, i. e., the potential of the model changes, tending to behave like an infinite well, then locating the fluctuations in a compact space. Because of this, to investigate compact configurations, we must modify the potential so that the kink mass can increase to larger and larger values. As proposed by Bazeia et. al. \cite{Bazeia}, one way to achieve this goal is to assume an adjustable potential that controls field fluctuations. Another possibility to study the influence of compact-like configurations on model dynamics and metrics is to assume that the vortex is governed by a known compact-like field \cite{ourPLB},

\begin{align}\label{eq3}
    \phi(r,\theta)=\nu \tanh{(r^\alpha)}\text{e}^{in\theta}, \hspace{1cm} \text{with} \hspace{1cm} \alpha=1,2,3,4,...
\end{align}
Here, $n$ denotes the winding number.

\begin{figure}[ht!]
\centering
\includegraphics[height=6.5cm,width=8cm]{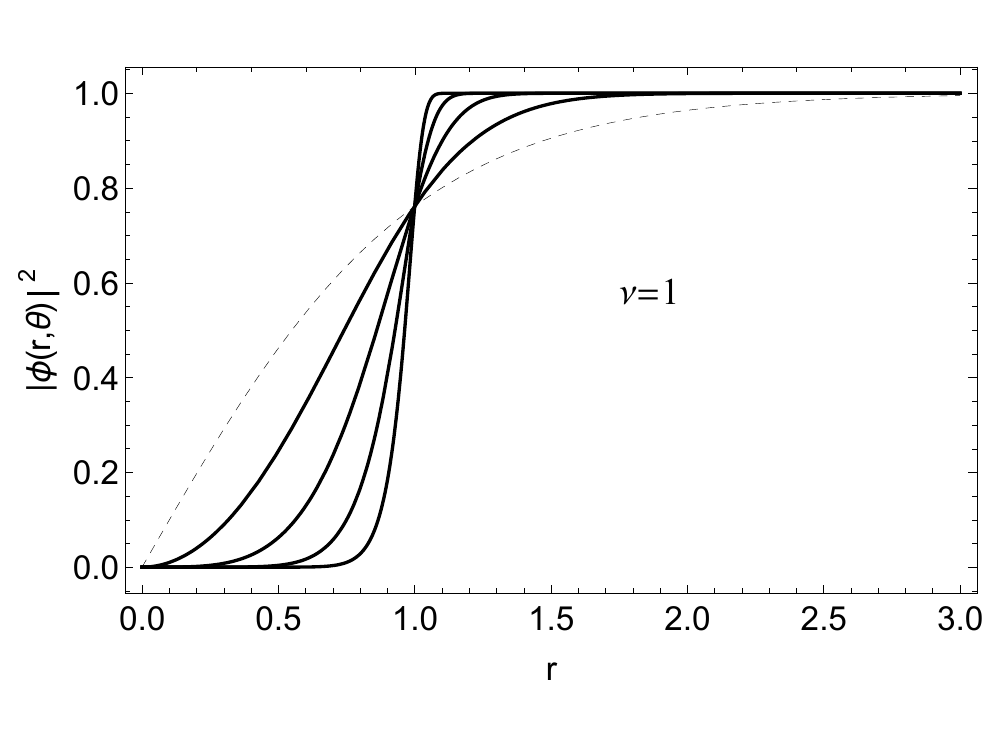}
\vspace{-0.5cm}
\caption{Profile of the matter field when $\alpha=1,2,4,8,10$. The dashed line describes the setting when $n=\alpha=1$.}
\label{fig1}
\end{figure}

It is interesting to mention that the profile chosen for the field of matter describes a topological structure. The compact-like profile of the matter field (\ref{eq3}) is clearly seen in Fig. \ref{fig1}.

To study the topological vortices of the model, consider rotationally symmetric static configurations so that the three-dimensional metric has the profile
\begin{align}\label{eq4}
    ds^2=-B(r)dt^2+\frac{1}{A(r)}dr^2+r^2d\theta^2.
\end{align}
This metric was considered in previous studies of Maxwell vortices in the AdS$_{3}$ background \cite{Edery,Albert}.

For the structures to be gauge invariant and rotationally symmetric, we assume that the gauge field is described by the ansatz:
\begin{align}\label{eq5} \textbf{A}_i(\textbf{r})=\varepsilon_{ij}\hat{x}^{j}\frac{a(r)}{er}.
\end{align}
For the previous ansatz, the vortices are purely magnetic and have a magnetic field given by
\begin{align}\label{eq6}
    F_{12}=-F_{21}=-\frac{1}{er}a'(r).
\end{align}
This profile of $F_{12}$ produces vortex with magnetic flux given by
\begin{align}\label{eq7}
    \Phi_B=\frac{2\pi}{e}[a(\infty)-a(0)].
\end{align}
Considering a topological boundary of the type $a(0)=0$ and $a(\infty)=n$, one arrives at
\begin{align}
    \Phi_B=\frac{2\pi n}{e}, 
\end{align}
with $n\in\mathbb{Z}$. Therefore, the structure has a quantized magnetic flux.

Considering the metric (\ref{eq4}) and the ansatz (\ref{eq3}) and (\ref{eq5}), we rewrite the Lagrangian density as
\begin{align}\label{eq8}\nonumber
    \mathcal{L}=& r\sqrt{\frac{B(r)}{A(r)}}\bigg[\frac{1}{16\pi G}(R-2\Lambda)+\frac{\nu^2\tanh(r^\alpha)^2(n-a(r))^2}{2r^2}+\frac{\nu^2\alpha^2 r^{2(\alpha-1)}\text{sech}(r^\alpha)^4 A(r)}{2}+\\
    -&\frac{A(r)a'(r)^2}{2e^2r^2}-\frac{\lambda}{4}(\nu^2\tanh{(r^\alpha)}^2-\nu^2)^2\bigg], 
\end{align}
where the notation line denotes the derivative with respect to the variable $r$. The Ricci tensor in terms of the metric functions $A(r)$ and $B(r)$ is
\begin{align}
    \label{eq9}
    R=-\frac{A'(r)}{r}+\frac{A(r)B'(r)^2}{2B(r)^2}-\frac{A(r)B''(r)}{B(r)}-\frac{A'(r)B'(r)}{2B(r)}-\frac{A(r)B'(r)}{r B(r)}. 
\end{align}

Note that the Lagrangian has three $r$ functions associated with compact-like vortex configurations, i. e., $A(r)$, $B(r)$, and $a(r)$. Their equations of motion are respectively:
\begin{align}
    \label{eq10} \nonumber
    &e^2 rA(r)B'(r)+B(r)[2e^2 r^2\Lambda+4\pi G(-2e^2\nu^2(n-a(r))^2\tanh(r^\alpha)^2+2e^2\nu^2r^{2\alpha}\alpha^{2} A(r)\text{sech}(r^{\alpha})^4+\\
    &+e^2r^2\lambda\nu^4\text{sech}(r^{\alpha})^4-2A(r)a'(r)^2)]=0,\\
    \label{eq11} \nonumber
    &-2e^2 r^2\Lambda+4\pi G[2e^2\nu^2(n-a(r))^2\tanh(r^\alpha)^2+2e^2\nu^2 r^{2\alpha}\alpha^{2}A(r)\text{sech}(r^\alpha)^4-e^2r^2\lambda\nu^4\text{sech}(r^\alpha)^4+\\
    &-2A(r)a'(r)^2]-e^2rA'(r)=0,\\
    \label{eq12} \nonumber
    &ra'(r)A(r)B'(r)+B(r)[2e^2 r\nu^2(a(r)-n)\tanh(r^\alpha)^2+a'(r)(-2A(r)+rA'(r))+\\
    &+2rA(r)a''(r)]=0.
\end{align}
The equations above describe the profile of the gauge field and the metric functions $A(r)$ and $B(r)$ that produce the solitonic profile of the matter field (see Eq. (\ref{eq3})).

\subsection{The case: $A(r)=B(r)$}

Before specifying our choice, let us mention that in Ref. \cite{Edery}, was studied the case $A(r)\neq B(r)$. In his study, Edery shows there are no black holes in this gravitational background. This remark is because to obtain a matter field with a solitary wave profile (kink), the solution of the metric functions is always positive-defined. We presume that a solitonic matter field (true soliton) produces the black hole. This hypothesis is supported by the fact that these structures interact with the geometry generating the emergence of such objects. So let us consider the simplest possible case, i. e., $A(r)=B(r)$. Moreover, in order to suppose the hypothesis of the arising of structures in (2+1)-dimensional, it is necessary to analyze some quantities. These quantities are the Ricci tensor, the quadratic invariant of the Ricci tensor, and the Kretschmann scalar to confirm that such structures are stable black holes. To begin our analysis, we write the Ricci tensor as

\begin{align}
    \label{eq13}
    R=-\frac{2A'(r)}{r}-A''(r).
\end{align}

By manipulating Eqs. (\ref{eq10}) and (\ref{eq11}), we have that
\begin{align}\label{eq14}
    \frac{B'(r)}{B(r)}=-\frac{A'(r)}{A(r)}+16\pi G\bigg(-\nu^2 r^{2\alpha-1}\alpha^2\text{sech}(r^{\alpha})^4+\frac{a'(r)^2}{e^2 r}\bigg).
\end{align}
Therefore, if $A(r)=B(r)$ the gauge field is given by 
\begin{align}\label{eq15}
    a'(r)=\pm e\nu r^{\alpha}\alpha\text{sech}(r^\alpha)^2,
\end{align}
i. e., 
\begin{align}\label{eq16}
    a(r)=a(0)\pm e\nu\alpha\int_{0}^{r}\, r^{_{'}\alpha}\text{sech}(r^{_{'}\alpha})^2\, dr'.
 \end{align}
For topological field structures, we assume that $a(0)=0$, thus we arrive at 
\begin{align}
    \label{eq17}
   a(r)=\pm e\nu\alpha\int_{0}^{r}\, r^{_{'}\alpha}\text{sech}(r^{_{'}\alpha})^2\, dr'.
\end{align}

By integration by part, we can rewrite $a(r)$ as
\begin{align}
    \label{eq18}
    a(r)=\pm e\nu\bigg[ r\tanh (r^\alpha)-\int_{0}^{r}\tanh(r^{_{'}\alpha})\, dr^{_{'}}\bigg].
\end{align}

Note that if $\alpha=1$ the gauge field takes its simplest possible form, namely,
\begin{align}
    \label{eq19}
    a(r)=\pm e\nu[r\tanh(r)-\text{ln}(\cosh(r))]. 
\end{align}

\begin{figure}[ht!]
\centering
\includegraphics[height=6cm,width=8cm]{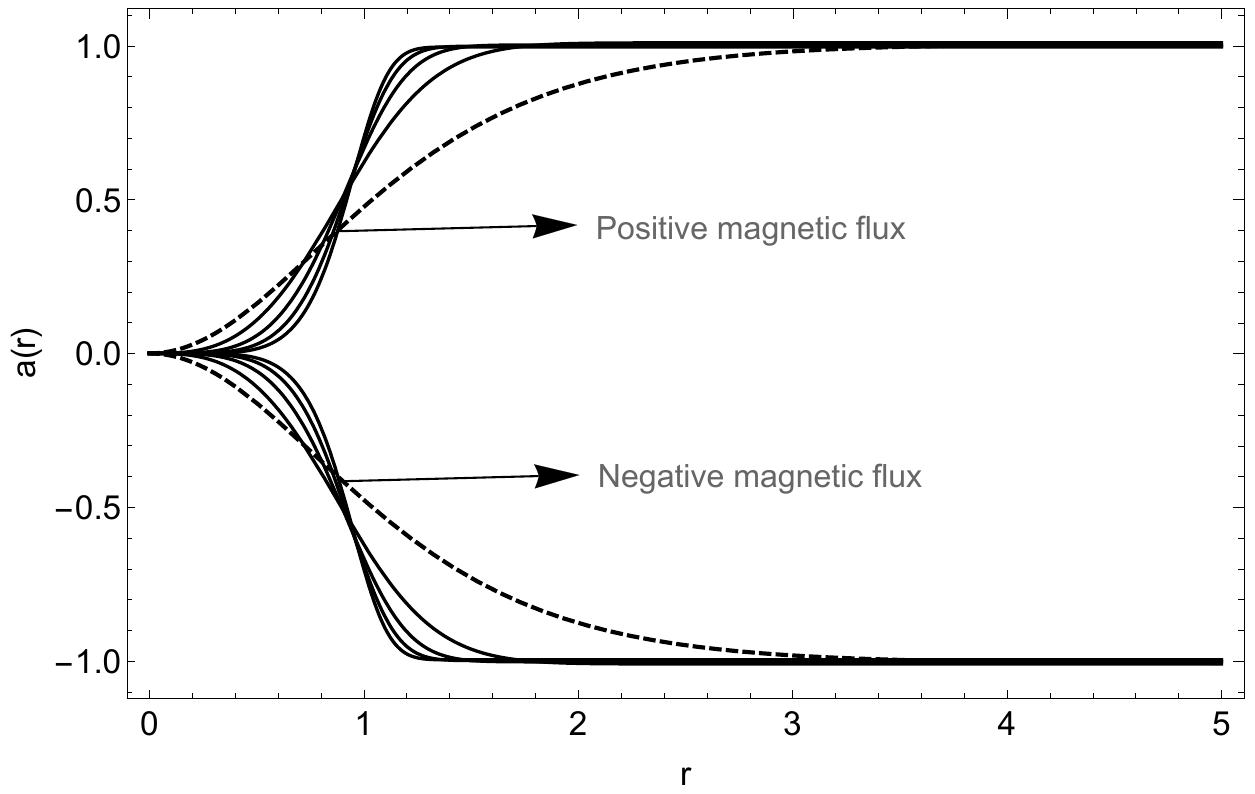}
\vspace{-0.2cm}
\caption{Gauge field associated with the matter field (\ref{eq3}). The dashed line describes the setting when $n=\alpha=1$.}
\label{fig2}
\end{figure}

For the cases of $\alpha>1$, it is convenient to calculate the behavior of $a(r)$ using some numerical technique. In our case, allow us to consider the differential equation (\ref{eq14}), and use a numerical integration in order to calculate the other gauge field profiles for $\alpha>1$. The numerical result of the gauge field is shown in Fig. \ref{fig2}.

Analyzing the equations of motion, we see that Eq. (\ref{eq12}) is not an independent equation of Eqs. (\ref{eq10}) and (\ref{eq11}). Thus, for the case $A(r)=B(r)$ we have only two independent differential equations. To study the metric functions, allow us to multiply Eq. (\ref{eq10}) by $B(r)$ and subtracting Eq. (\ref{eq9}) from Eq. (\ref{eq10}). Then, we have
\begin{align}
\label{eq20}
A'(r)=B'(r)=-2\pi\Lambda+2\pi G\bigg[\frac{4\nu^2(n-a(r))^2\tanh (r^\alpha)^2}{r}-2r\lambda\nu^4\text{sech}(r^\alpha)^{4}\bigg].
\end{align}

\begin{figure}[ht!]
\centering
\includegraphics[height=6.5cm,width=8cm]{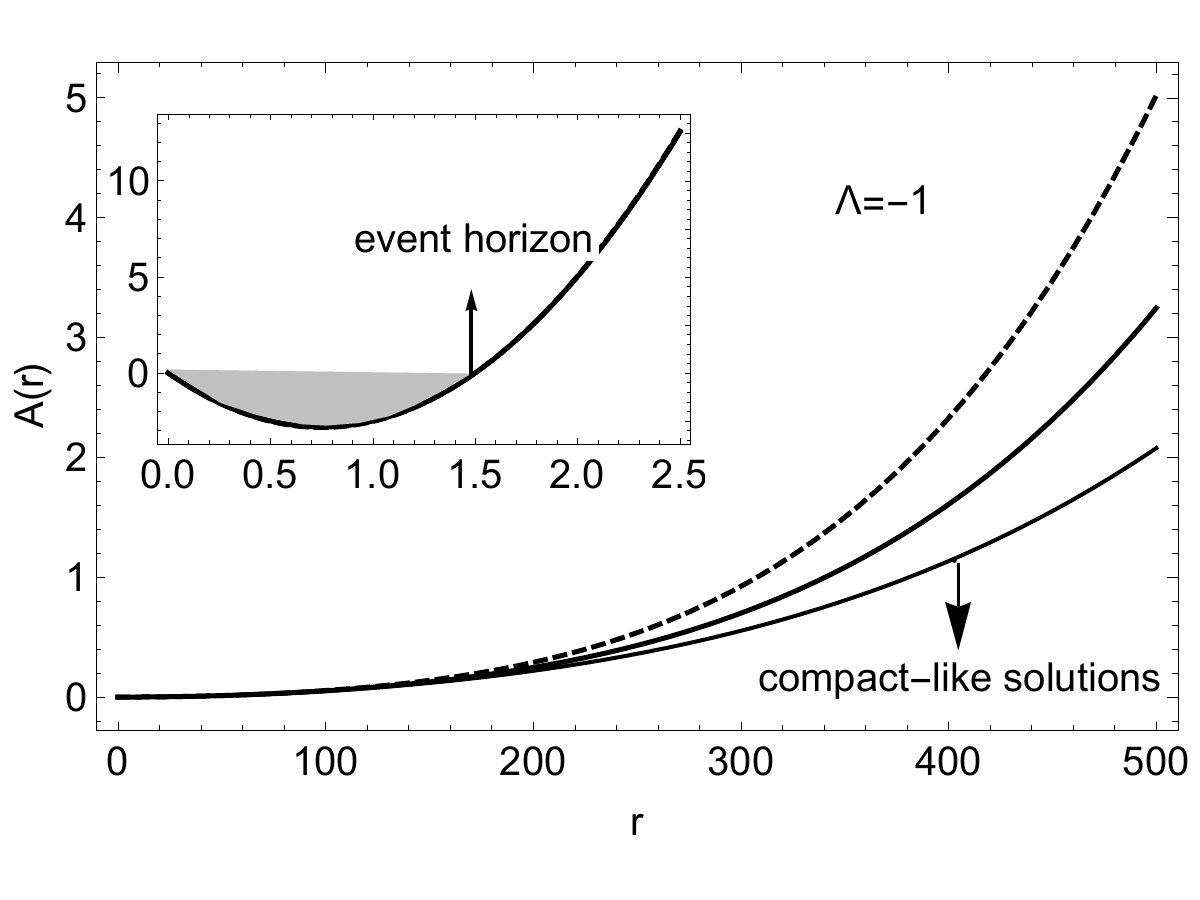}
\includegraphics[height=6.5cm,width=8cm]{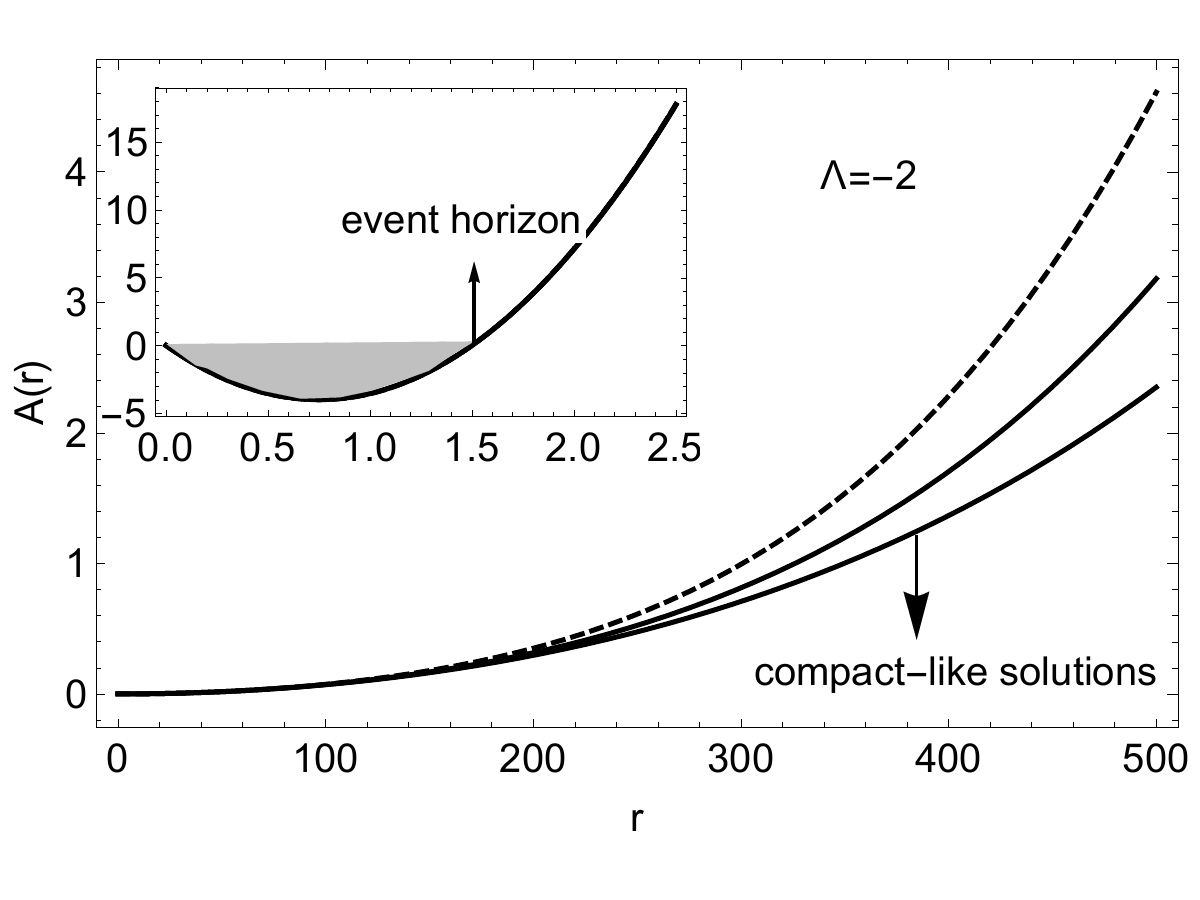}\\
\vspace{-0.2cm}
\hspace{0.5cm} (a) \hspace{7.5cm} (b)
\vspace{-0cm}
\caption{The numerical result of the metric functions when $n=\alpha=\nu=\lambda=1$ and $G=1/4\pi$. In all the cases, one assumes the integration constant ($C_0$) is one. (a) Plot when $\Lambda=-1$. (b) Plot when $\Lambda=-2$.}
\label{fig3}
\end{figure}

\begin{figure}[ht!]
\centering
\includegraphics[height=6.5cm,width=8cm]{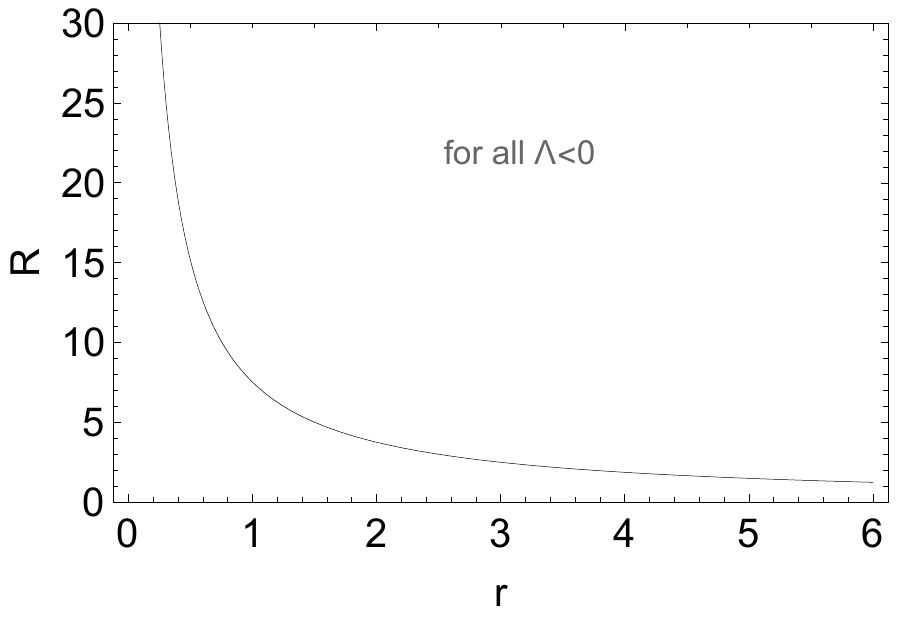}
\vspace{-0.2cm}
\caption{Curvature scalar independent of the $\alpha$ parameter when $n=\alpha=\nu=\lambda=C_0=1$ and $G=1/4\pi$.}
\label{fig4}
\end{figure}

The numerical solution of Eq. (\ref{eq20}) is shown in Fig. \ref{fig3}. Note that the metric function $A(r)$ in the vicinity of the vortex sweeps a negative domain and the distance from the vortex becomes fully positive, so there is an event horizon around the value $r=1.5$. It is notable from the numerical simulation that the metric function $A(r)$ near the origin and therefore at the core of the vortex has a polynomial profile \footnote{One can verify this statement by performing an analytical analysis around the origin, i.e., $r=0$. In this case, we obtain $\lim_{r\to 0}A(r)\simeq -\Lambda r^2+\mathcal{O}(r^2)$.}. On the other hand, away from the vortex, it has an asymptotic behavior of type $r^{2}$ \footnote{One verifies this analysis in Sec. 2.2.1.}. From Figs. \ref{fig3}(a) and \ref{fig3}(b), it is observed that regardless of the cosmological constant, if the field variable becomes compact-like the metric functions tend to behave asymptotically linearly.  In other words, for compact-like configurations ``kink mass'' increases and this means that for the confinement of the field fluctuations to occur, the metric function has to tend to a linear behavior, and therefore, the Higgs potential confines the compacted structures.  Note that the result found in Ref. \cite{Edery}, is obtained if we consider a field configuration similar to kink (however, quite located). An interesting result is that for all matter field profiles, we will have a black hole located at the origin of the system. The curvature scalar associated with the metric is shown in Fig. \ref{fig4}.

As mentioned at the beginning of the section, it is necessary to analyze some quantities to analyze the stability of the black holes. Using the numerical solutions of the metric functions, we will investigate the quadratic invariant of the Ricci tensor and the Kretschmann scalar.

\begin{figure}[ht!]
\centering
\includegraphics[height=6.5cm,width=8cm]{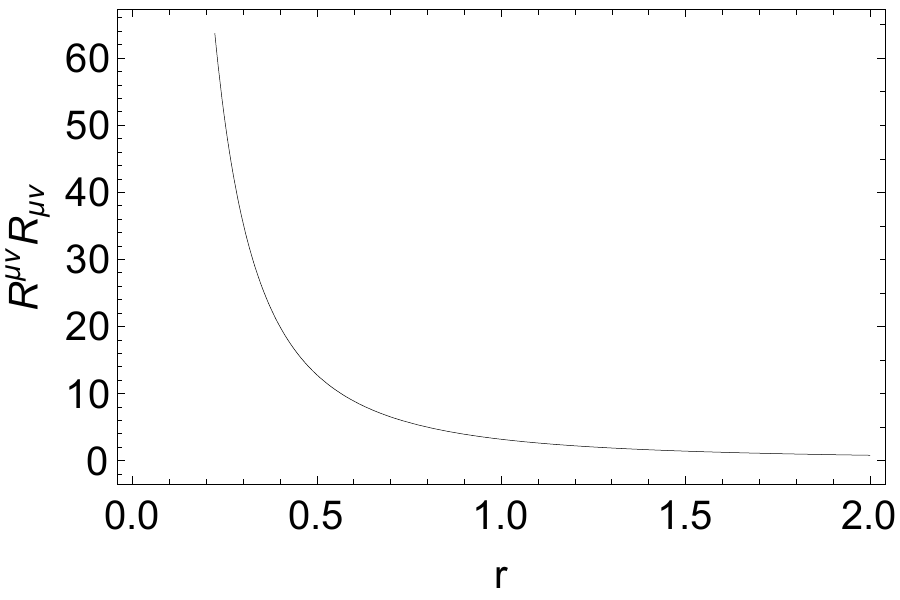}
\includegraphics[height=6.5cm,width=8cm]{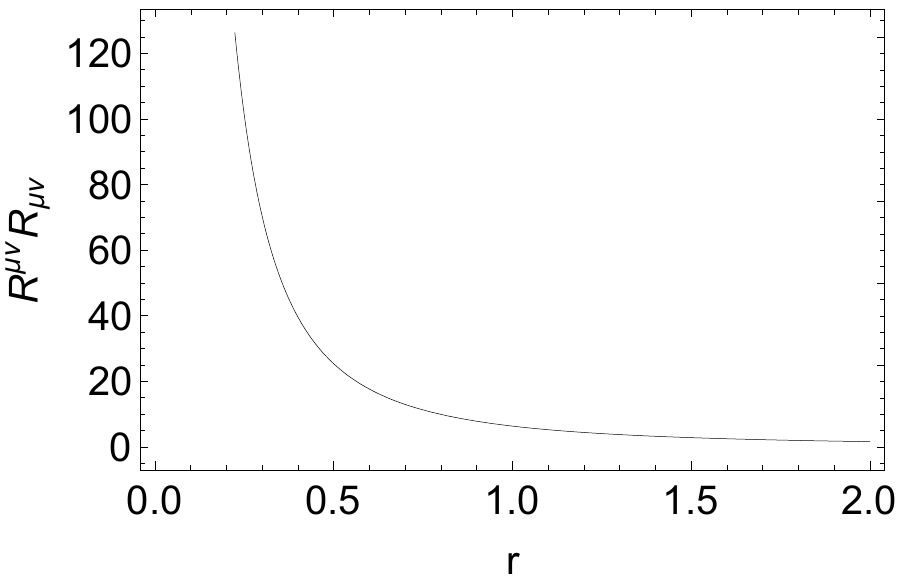}\\
\vspace{-0.2cm}
\hspace{0.5cm} (a) \hspace{7.5cm} (b)
\vspace{-0cm}
\caption{The quadratic invariant of the Ricci tensor when $n=\alpha=\nu=\lambda=C_0=1$ and $G=1/4\pi$. (a) Plot when $\Lambda=-1$. (b) Plot when $\Lambda=-2$.}
\label{fig5}
\end{figure}

By direct calculation for the metric (\ref{eq4}), the quadratic invariant of the Ricci tensor is given by
\begin{align}
    R^{\mu\nu}R_{\mu\nu}=\frac{1}{2r^2}A'(r)[A'(r)+r A''(r)].
\end{align}
The numerical result of the Ricci quadratic invariant is displayed in Fig. \ref{fig5}. Note that the quadratic invariant of the Ricci tensor has a singularity at the center of the black hole vortex and a constant behavior away from that structure. 

In addition, the Kretschmann scalar is defined as
\begin{align}\label{kretschmann}
K= R^{\mu\nu\tau\sigma}R_{\mu\nu\tau\sigma},
\end{align}
where $R_{\mu\nu\tau\sigma}$ is the Riemann curvature tensor. For our structures, the Kretschmann scalar is reduced to
\begin{align}\label{Kretschmann}
K= R^{\mu\nu\tau\sigma}R_{\mu\nu\tau\sigma}=\frac{A'(r)^2}{r^2}.
\end{align}

\begin{figure}[ht!]
\centering
\includegraphics[height=6.5cm,width=8cm]{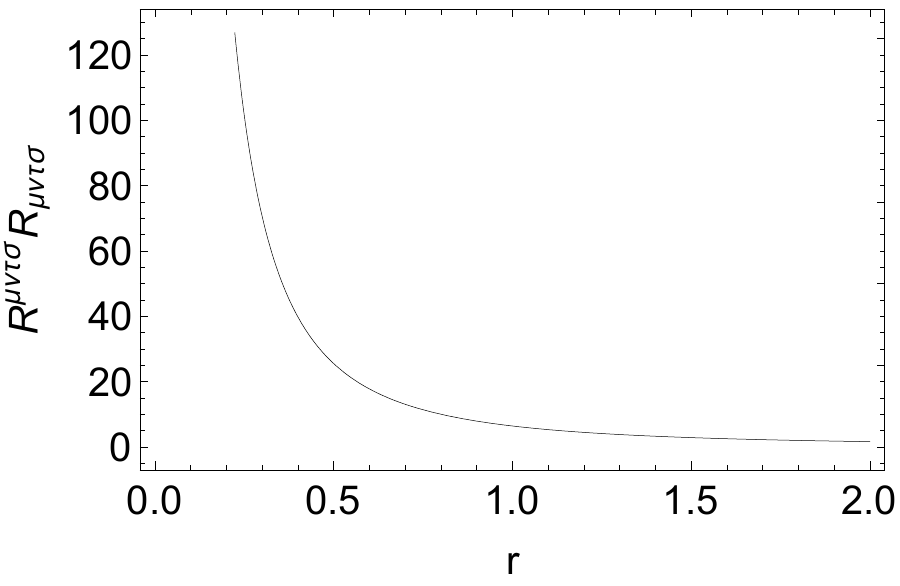}
\includegraphics[height=6.5cm,width=8cm]{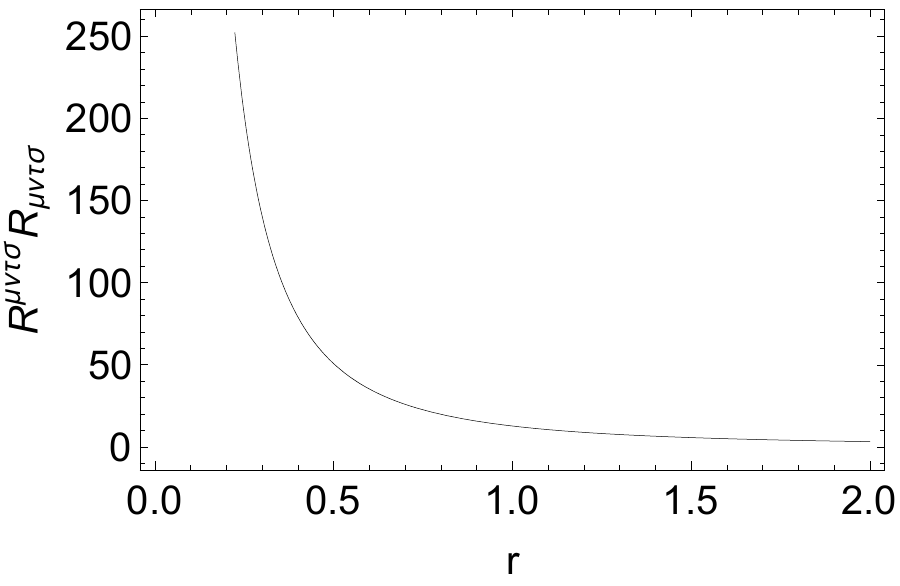}\\
\vspace{-0.2cm}
\hspace{0.5cm} (a) \hspace{7.5cm} (b)
\vspace{-0cm}
\caption{Kretschmann scalar  when $n=\alpha=\nu=\lambda=C_0=1$ and $G=1/4\pi$. (a) Plot when $\Lambda=-1$. (b) Plot when $\Lambda=-2$.}
\label{fig6}
\end{figure}

By the Kretschmann scalar (Fig. \ref{fig6}), it is observed that there is a singularity at the origin, and asymptotically the gravitational field is described only by the cosmological constant. Thus, these results [see Fig. \ref{fig5} and \ref{fig6}] suggest the existence of a black hole located at $r=0$.

\subsection{The asymptotic behavior}

 \subsubsection{The metric function}
 
Let us investigate the asymptotic behaviors of metric functions. For this purpose, one can analytically solve Eq. (\ref{eq10}), in the VEV of the theory, i. e., $a\to n$ and $\vert\phi\vert\to \nu$. This gives us the equation for $A(r)$, namely,
 \begin{align}\label{eq21}
     A'(r)=-2r\Lambda,
 \end{align}
 i. e.,
 \begin{align}\label{eq22}
     A_{0}(r)=-\Lambda r^2+C. 
 \end{align}
 The subscript `$0$' describes the vacuum state. $C$ is an integration constant that defines the initial conditions at the source. In addition to an $AdS_3$ model, we are interested in studying an asymptotically Minkowski spacetime ($\Lambda=0$). For this, we must have $C=1$.

 Meanwhile, in the asymptotic limit, i.e., in $r\to \infty$, one obtains
 \begin{align}\label{eq23}
     A(r)=-\Lambda r^2+D.
 \end{align}
Here the constant $D$ differs from the constant $C$.
 
 \subsubsection{The gauge field}
 
The analysis of the asymptotic behavior of gauge fields is convenient to understand the topological structure in more detail. Analyzing the profile of the gauge field of Eq. (\ref{eq14}) near the origin, we get that for $r\ll 1$, the gauge field has a profile
 \begin{align}\label{eq24}
    &a(r)\simeq \pm e\nu\bigg(\frac{1}{2}r^2-\frac{1}{3}r^3+\mathcal{O}(r^5)\bigg), \, \, \, \, \, \text{if} \, \, \, \, \, \alpha=1.
 \end{align}
The gauge field has a profile of an ascending monotonic function near the origin. This causes the greatest contribution of the magnetic flux to be close to the vortex core. Note that, in our theory, we have a purely magnetic black hole-like vortex structure. This allows us to assume that the thermodynamics describing these structures must be similar to black hole thermodynamics. For more details on the thermodynamics of these structures, see Sec. IV.

\section{The ADM mass}

Proposed by Arnowitt-Deser-Misner (ADM) in 1959 \cite{Arnowitt}, the ADM formalism is a Hamiltonian formulation of general relativity. This formalism plays a relevant role in quantum gravity theory and numerical relativity \cite{Shafer,Shafer1}. Indeed, ADM formalism presents itself as an approach to studying the theoretical content of fields rather than geometric information. ADM formalism gives us information about ADM energy (ADM mass). Naturally, this occurs because the gravitational mass (total energy) in the asymptotically-flat spacetime relates to the gravitational Hamiltonian of the theory \cite{Poisson}. In this formalism, the concept of ADM mass informs us about measurements of the field asymptotically flat. Furthermore, it is possible to find discussed in Ref. \cite{Abbott} the extension of the ADM formalism to asymptotically AdS spacetime.

For the analysis of the ADM mass, the model is asymptotically embedded in an AdS$_{3}$ spacetime. The Killing vector is timelike, since no metric components are time-dependent. The notion of conserved energy is applied to matter embedded in AdS$_3$. To calculate the ADM mass, we use its generalized form for $2+1$-dimensions \cite{Poisson}, namely,
\begin{align}\label{eq25}
    M=-2\alpha_0\lim_{S_t\to R}\oint_{S_t}(k-k_0)N(R)\sqrt{\sigma}d^2\theta.
\end{align}
Above, $S_t$ is a circular contour at spatial infinity (i. e. $r=R$), $\sigma_{AB}$ is the metric at $S_t$, and $k=\sigma^{AB}k_{AB}$ is the extrinsic curvature of $S_t$ embedded in the two-dimensional spatial surface obtained by defining $t$ as constant. Similarly, $k_0$ is the extrinsic curvature of $S_t$ embedded in the two-dimensional spatial surface of AdS$_3$. As shown in Ref. \cite{Edery}, for the metric used, we have
\begin{align}\label{eq26}
    k=\frac{\sqrt{A(r)}}{R}, \hspace{1cm} k_{0}=\frac{\sqrt{A_0(R)}}{R}, \hspace{1cm} N(r)=\sqrt{A_0(R)}, \hspace{1cm} \sqrt{\sigma}=R.
\end{align}

By replacing Eq. (\ref{eq25}) in Eq. (\ref{eq24}), we have
\begin{align}\label{eq27}
    M=&4\pi\alpha_0\{A_0(R)-[A_0(r)A(R)]^{1/2}\}.
\end{align}
It is interesting to highlight that ADM mass (\ref{eq27}) in the static case is of great physical importance. Principally, this relevance is because an isolated body can emit radiation, and the rate of mass change is related to the external flux of radiated energy \cite{Poisson}.

The mass expression is simplified by noting that
\begin{align}\label{eq28}
    A(R)=-R^2\Lambda+C+(D-C)=A_0(R)+(D-C).
\end{align}

Note that for $A_0(R)=A(R)$, the ADM mass is null. However, when $R$ is large, the binomial expansion reduces the mass expression to
\begin{align}\label{eq29}
    M=2\pi\alpha_0(C-D).
\end{align}
As a matter of fact, the expression obtained is constant. The constant mass is an indication that spacetime is radiant. Indeed, the ADM mass (\ref{eq25}) represents all the energy (mass) contained in the two-dimensional spatial surface. In this case, we have that this general quantity is constant. This constancy is because the surface intercepts the black hole vortex, whose mass increases (increases radiation) when $\Lambda$ decreases (as we will see later). On the other hand, the surface will also intercept the radiation. Thus, the ADM mass is responsible for both forms of energy, and in this case we have a conserved quantity.

Therefore, the mass expression in the AdS$_3$ background is
\begin{align}\label{eq30}
    M_{AdS_{3}}=2\pi\alpha_0 (A_0(R)-A(R)).
\end{align}
The spacetime from the BTZ black hole is asymptotically retrieved. In fact, for BTZ models, we define $\alpha_0$ as $1/2\pi$ and $C=0$. In this case, $M=-D$, so that $A(R)= -R^2\Lambda-M$, and $B(R)=\kappa A(R)$ represents the spacetime of the black hole BTZ asymptotically. For more details see Ref. \cite{Edery}. 

\subsection{Integral representation of the mass}

We can solve the equations of motion for the metric function $A(r)$ in terms of matter fields. To do this, let us consider Eq. (\ref{eq20}) for $A'(r)$. The integral representation for the ADM mass of the vortex embedded in the AdS$_3$ background is
\begin{align} \label{eq31}
M_{\text{AdS}_3}=&2\pi\alpha_0\int\, \bigg\{-2\pi\Lambda+2\pi G\bigg[\frac{4\nu^2(n-a(r))^2\tanh (r^\alpha)^2}{r}-2r\lambda\nu^4\text{sech}(r^\alpha)^{4}\bigg]\bigg\}\, dr,
\end{align}
where the mass $M_{\text{AdS}_3} $ does not depend purely on matter fields. Indeed, the mass is dependent on the cosmological constant and the gravitational constant when the matter field is described by a topological soliton. The numerical value of the mass AdS$_3$ is shown in Table \ref{tab1}, for various values of the cosmological constant and for the first values of $\alpha$.

\begin{table}[ht!]
\centering
\caption{Table with the numerical results of the mass $M_{AdS_{3}}$ with $R=10$ and $G=1$.}
\label{tab1}
\resizebox{5.5cm}{5cm}{%
\begin{tabular}{|c|c|c|}\hline\hline
$\alpha$ & (n,$\nu$, $\Lambda$) & M$_{AdS_3}$  \\ \hline\hline
\multirow{3}{*}{1}  & $(1,1,-1)$ & 545.40  \\ \cline{2-3}
  & $(1,1,-2)$ &  940.40 \\ \cline{2-3}
  & $(1,1,-3)$ &  1334.97 \\ \hline\hline
  \multirow{3}{*}{2} & $(1,1,-1)$ &  482.06 \\ \cline{2-3}
  & $(1,1,-2)$ &  876.84 \\ \cline{2-3}
  & $(1,1,-3)$ &  1271.63 \\ \hline
\end{tabular}}
\end{table}

\section{The thermodynamics of black hole vortex}

Let us use the Hamilton-Jacobi formalism to study black hole thermodynamics through the tunneling approach \cite{Srinivasan1998ty,Angheben2005rm,Kerner2006vu,Mitra2006qa,Akhmedov2006pg}. The main purpose of this technique is to obtain the Hawking temperature for the black hole generated by the profile of the metric function (\ref{eq20}).

The basic idea of the tunneling method is to calculate the probability of particles created near the event horizon escaping the black hole through quantum tunneling. This is possible when we interpret Hawking radiation as an emission process through the black hole. This emission process occurs due to the spontaneous creation of pairs of particles within the event horizon. Particles with negative energy remain inside the black hole and contribute to its mass loss. The positive energy particle manages to escape the horizon through a ``tunnel'' towards infinity. So it is possible to associate the tunneling probability to the black hole temperature. The main advantage of using the tunneling method to investigate the thermodynamics of black hole is due the thermodynamic properties are related to the geometry of spacetime, allowing a wide application to spacetime varieties \cite{Gomes2018oyd,Jiang2006,Kerner2007rr,Ma2014qma,Maluf2018lyu,Gomes2020kyj}.

Near the black hole horizon, we have only the temporal and radial terms of the metric once the angular part is red-shifted away. Thus, the metric becomes 2-dimensional, i.e., 
\begin{align}
ds^2=-A(r)dt^2+A(r)^{-1}dr^2.
\end{align}
One can see more details on this in Refs. \cite{SilvaBrito,GMA}.

Let us consider a perturbation of a massive scalar field $\phi$ around the black hole background, i. e.,
 \begin{align}
     \hbar^2g^{\mu\nu}\nabla_\mu\nabla_\nu\phi-m^2\phi=0.
 \end{align}
The above equation is a kind of Klein-Gordon equation where $m$ is the mass associated with the $\phi$ field. Using the decomposition on spherical harmonics, one arrives at
 \begin{align}\label{0101}
    -\partial_t^2\phi+A(r)^2\partial_r^2\phi+\frac{1}{2}\partial_rA(r)^2\partial_r\phi-\frac{m^2}{\hbar^2}A(r)\phi=0.
 \end{align}

Since $\phi$ is a semiclassical field associated with particles created in the black hole, we can use the so-called WKB approximation \cite{Sakurai} through the ansatz \cite{Srinivasan1998ty,Angheben2005rm,Kerner2006vu,Mitra2006qa,Akhmedov2006pg}
 \begin{align}
     \phi(t,r)=\exp\Big[\frac{1}{\hbar}\mathcal{T}(t,r)\Big],
 \end{align}
to obtain the solution of Eq. (\ref{0101}).

Note that the Eq.(\ref{0101}) for the lowest order in $\hbar$ is
 \begin{align}\label{0102}
 (\partial_t\mathcal{T})^2-A(r)^2(\partial_r\mathcal{T})^2-m^2A(r)\phi=0,
 \end{align}
where particle-like solutions have the form \cite{Srinivasan1998ty,Angheben2005rm,Kerner2006vu,Mitra2006qa,Akhmedov2006pg}
\begin{align}\label{0103}
\mathcal{T}(t,r)=-\omega t+W(r).
\end{align}
Here, $\omega$ is a constant that can be thought as the energy of the emitted radiation. By replacing the solution (\ref{0103}) in Eq. (\ref{0102}), it is found that
\begin{align}\label{0104}
W(r)=\pm\int{\frac{dr}{A(r)}\sqrt{\omega^2-m^2A(r)}},
\end{align}
where the positive solution represents the output particle, and the negative solution the input particle. Let us focus on the output solution, as they represent particles emitting radiation as they cross the event horizon. For simplicity, it is convenient to assume the approximation of the function $A(r)$ near the event horizon, i. e., $r_+$, in the form
\begin{align}
A(r)=A(r_+)+A'(r_+)(r-r_+)+...
\end{align}

In this way, Eq. (\ref{0104}) then becomes
\begin{align}\label{0105}
W(r)=\int{\frac{dr}{A'(r_+)}\frac{\sqrt{\omega^2-m^2A'(r_+)(r-r_+)}}{(r-r_+)}}.
\end{align}

Using the residue theorem to solve the integral (\ref{0105}), we obtain that
\begin{align}
W(r_{+})=\frac{2\pi i \omega}{A'(r_{+})}+(\text{real contribution}).
\end{align}

Therefore, the probability of a particle escaping the black hole through tunneling is \cite{Srinivasan1998ty,Angheben2005rm,Kerner2006vu,Mitra2006qa,Akhmedov2006pg} 
\begin{align}\label{0106}
\Gamma\sim\exp(-2im\mathcal{T})=\exp\Big[-\frac{4\pi\omega}{A'(r_+)}\Big].
\end{align}

By comparing the tunneling probability (\ref{0106}) with the Boltzmann factor $e^{-\omega/T}$, it is directly concluded that the Hawking temperature is
\begin{align}\label{0107}
T_H=\frac{\omega}{2im\mathcal{T}}=\frac{A'(r_+)}{4\pi}.
\end{align}

\begin{figure}[ht!]
\centering
\includegraphics[height=6.5cm,width=8cm]{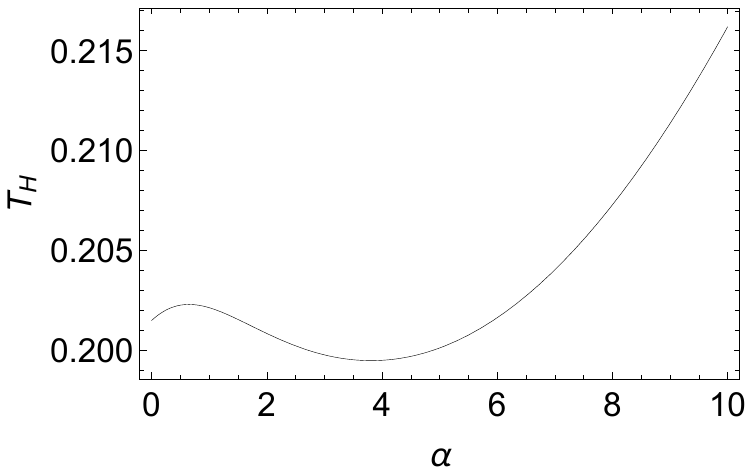}
\includegraphics[height=6.5cm,width=8cm]{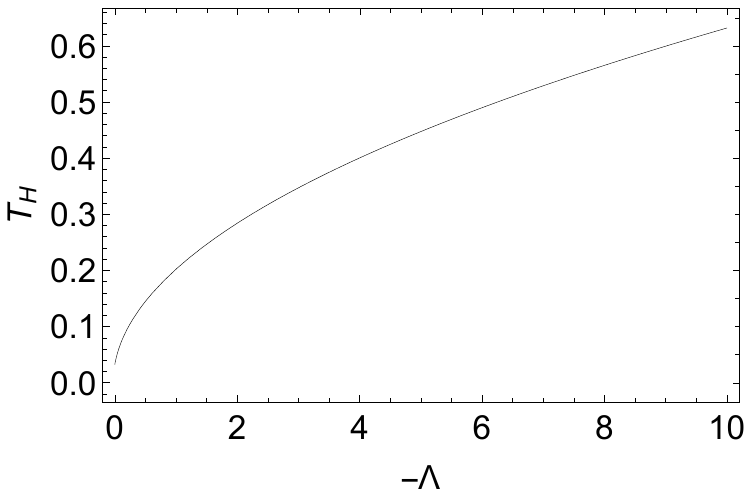}\\
\vspace{-0.2cm}
\hspace{0.5cm} (a) \hspace{7.5cm} (b)
\vspace{-0cm}
\caption{Numerical result of the Bekenstein-Hawking Temperature. (a) Plot when $\Lambda=-1$ and $\lambda=\nu=1$. (b) Plot when $\lambda=\nu=\alpha=1$.}
\label{figT}
\end{figure}

By analyzing the numerical results presented in Sec. I, it can be seen that $r_+\approx 1.5$. Note that the Bekenstein-Hawking temperature $T_H$ can only be obtained numerically for this model. For the case $\lambda=\nu=\alpha=1$ and $\Lambda=-1$, $T_H\approx 0.2021$. The figure \ref{figT} represents the Bekenstein-Hawking temperature, varying the parameter $\alpha$ that controls the formation of the compact-like structure of the model (Fig.\ref{figT} $a$), and varying the cosmological constant $ \Lambda$ (Fig.\ref{figT} $b$).

\section{Final remarks}

In this work, the influence of the matter field and gauge field on the metric functions of an AdS$_3$ space is studied. For this study, an axially symmetrical matter field was considered, with a field variable described by a kink capable of becoming compacted. For simplicity, we considered for our study a particular case of the 3D metric ($A(r)=B(r)$). The study point that the two-dimensional structures of the space AdS$_3$ are topological and magnetic black hole vortices have a quantized flux.

When studying the particular case $A(r)=B(r)$, an interesting result emerges due to the profile of the matter field, namely, the metric functions must be type black hole with the horizon in $r_{+}\approx 1.5$. This result is interesting because it generalizes the result of Ref. \cite{Edery}. In that work,  a variable field kink type generates positive-defined metric functions with a singularity in $r=0$, and therefore the structure will not be a black hole vortex. We believe that black hole vortices can arise in the model presented in Ref. \cite{Edery}, if some adjustment mechanism of the matter field is used to adjust the field variable and make it a strictly pure kink.

Our black hole vortex results are further confirmed by the Kretschmann and Ricci scalar analysis, as well as the model curvature. Analyzing these quantities, it can be seen that asymptotically the metric solutions should depend only on the cosmological constant. Consequently, this cosmological constant will directly influence the numerical results of the mass M$_{AdS_3}$.

By inspection, it was noted that the cosmological constant must be negative or zero, implying the arise of black hole vortices, These structures have already been shown to exist in the so called Bergshoeff, Hohm and Townsend (BHT) massive gravity \cite{Oliva}, which also occurs in (2+1)-dimension. Therefore, this suggests that our results are compatible with the existing literature.

Through tunneling formalism, it was shown that the thermodynamic properties of the black hole are modified by the cosmological constant, and by the parameter of compactification of field. The black hole temperature change occurs due to the modification of superficial gravity, which is a consequence of the geometry modification. Changing the geometry is also responsible for changing the surface area of the black hole vortex. It is observed that the more compact structures of the matter field will influence the geometry so that the temperature of the black hole vortices will be higher.

Finally, it is observed that the compactification of the matter field changes the profile of the gauge field, making the magnetic field of the vortex more intense at the vicinity of the core, and with the emission of quantized magnetic flux. Compact-like topological configurations also generate structures with lower ADM mass, when compared to structures governed by the field variable with the profile of a kink. Note that despite the compact-like structures having a smaller mass profile, they still have a very high contribution when compared with the mass result of Ref. \cite{Edery}. We believe that this feature is due to the explicit contribution of the cosmological constant in mass, and due to the fact that the model admits black holes, which was not possible in Ref. \cite{Edery}.

An continuation of this study could be to consider the generalizations of the model, as well as the study of the influence of Lorentz violation on this theory. We hope to carry out this study in future work.

\section*{Acknowledgments}

C. A. S. Almeida thanks to Conselho Nacional de Desenvolvimento Cient\'{\i}fico e Tecnol\'{o}gico (CNPq), n$\textsuperscript{\underline{\scriptsize o}}$ 309553/2021-0. F. C. E. Lima is grateful to Coordena\c{c}\~{a}o de Aperfei\c{c}oamento de Pessoal de N\'{i}vel Superior (CAPES), n$\textsuperscript{\underline{\scriptsize o}}$ 88887.372425/2019-00. The authors thank the referee for his valuable review.


\end{document}